\documentclass[onecolumn,superscriptaddress, aps,pra ]{revtex4-2}
\makeatletter
\newcommand*{\rom}[1]{\expandafter\@slowromancap\romannumeral #1@}
\makeatother
\usepackage{color}
\usepackage{graphicx}
\usepackage{hyperref}
\usepackage{amsmath,amssymb}
\usepackage{epstopdf}
\usepackage{subcaption}
\graphicspath{{figs/}}

\def\beq{\begin{equation}}
\def\eeq{\end{equation}}
\def\bea{\begin{eqnarray}}
\def\eea{\end{eqnarray}}

\begin{document}
\title{\boldmath Rotation curves of dwarf galaxies and dark matter as a condensate of $q$-deformed bosons: reconstruction and fits of the deformation profile}
\author {Mahnaz Maleki}
\email{m.maleki@uma.ac.ir}
\affiliation{Department of Physics, University of Mohaghegh Ardabili, P.O. Box 179, Ardabil, Iran}
\author {Asef Kheirandish}
\affiliation{Department of Physics, University of Mohaghegh Ardabili, P.O. Box 179, Ardabil, Iran}
\author {Morteza Nattagh Najafi}
\affiliation{Department of Physics, University of Mohaghegh Ardabili, P.O. Box 179, Ardabil, Iran}
\author {Hosein Mohammadzadeh}
\email{mohammadzadeh@uma.ac.ir}
\affiliation{Department of Physics, University of Mohaghegh Ardabili, P.O. Box 179, Ardabil, Iran}
\pacs{}
\begin{abstract}
Dark-matter-dominated dwarf galaxies are a sensitive testing ground for models in which dark matter is a self-interacting Bose-Einstein condensate (BEC). In the Thomas-Fermi limit the standard BEC halo has the density profile $\rho\propto\sin(kr)/(kr)$, so that the shape of its rotation curve is fixed up to two scale parameters. Here we consider a condensate of $q$-deformed bosons. We show that, in the Thomas-Fermi limit, the deformation enters the structure of the halo only through the chemical potential of the condensate; consequently, a spatially uniform deformation parameter leaves the standard BEC profile unchanged. A nontrivial effect arises only if the deformation is allowed to vary across the halo, $q\to q(r)$. In that case the density obeys the modified equation $\nabla^{2}(\rho+\chi)+k^{2}\rho=0$, where $\chi(r)\propto-\ln q(r)$ has the dimension of a mass density and plays the role of a position-dependent shift of the chemical potential. We invert this equation to express $\chi$ directly in terms of the rotation curve, and we reconstruct the deformation profile of eight dwarf galaxies (IC~2574, NGC~2366, Holmberg~I, Holmberg~II, M81~dwB, DDO~53, DDO~39, and DDO~154). In every galaxy $\chi$ is depressed in the inner region, $kr\lesssim1$, relative to the outer halo, where it is approximately constant in most cases: the standard condensate describes the outer halo, while the core requires a weaker deformation, i.e., $q$ closer to unity. The profiles share this qualitative shape, but neither the depth nor the width of the depression is universal. Modelling the depression by a Gaussian, with two additional parameters, we fit the rotation curves with rms residuals of $0.6$--$2.6$~km\,s$^{-1}$, against $1.6$--$8.5$~km\,s$^{-1}$ for the standard condensate. According to the Bayesian information criterion, the deformed condensate is preferred over the standard condensate, the Navarro-Frenk-White profile, and the Burkert profile in four galaxies, is indistinguishable from the best of them in two, and is disfavoured in two. We discuss the limitations of these results, in particular the absence of error bars in the data, the neglect of baryons, and the phenomenological status of a position-dependent deformation.

\textbf{keywords}: dark matter theory, rotation curves of galaxies, dwarfs galaxies
\end{abstract}
\maketitle
\section{Introduction}\label{1}
Dark matter is one of the fundamental components of the standard cosmological picture and plays a central role in the formation and evolution of cosmic structures. Although it does not couple to electromagnetic radiation in the way ordinary baryonic matter does, its gravitational effects are evident on a wide range of scales, from galaxies and galaxy clusters to the large-scale structure of the Universe. The $\Lambda$CDM model describes many cosmological observations successfully, but the microscopic nature of dark matter remains unknown, and its distribution and dynamics on galactic scales continue to be an open problem \cite{peebles2015physical,weinberg2008cosmology,trimble1987existence}.

A well-known difficulty of the cold dark matter scenario is the discrepancy between the central density profiles obtained in numerical simulations and those inferred from observations. Dark-matter-only simulations predict centrally concentrated haloes such as the Navarro-Frenk-White (NFW) profile \cite{navarro1997universal},
\begin{equation}
\rho_{\rm NFW}(r)=\frac{\rho_s}{(r/r_s)(1+r/r_s)^2},
\end{equation}
where $\rho_s$ and $r_s$ are the characteristic density and the scale radius. Observations of several dark-matter-dominated dwarf and low-surface-brightness galaxies indicate instead shallower, cored central distributions. This is the core-cusp problem, and it has motivated alternative descriptions of the halo. Dwarf galaxies are particularly useful in this respect because their dynamics is dominated by the dark component, so that their rotation curves constrain the halo density profile rather directly, in particular in the inner region where different models differ most.

One alternative is provided by models in which dark matter particles are bosons that form a Bose-Einstein condensate (BEC). The halo is then described by a condensate wave function obeying the Gross-Pitaevskii equation (GPE) coupled to the Poisson equation, and the repulsive self-interaction of the condensate naturally produces a nonsingular density distribution with a central core. This scenario was studied in detail by Harko \cite{harko2011bose}, who modelled dark matter as a strongly coupled dilute condensate in the Thomas-Fermi (TF) approximation and compared the resulting rotation curve with those of eight dwarf galaxies: IC~2574, NGC~2366, Holmberg~I, Holmberg~II, M81~dwB, DDO~53, DDO~39, and DDO~154. The TF halo has the density profile $\rho=\rho_c\sin(kr)/(kr)$, which contains only two parameters, the central density $\rho_c$ and the halo radius $\pi/k$. The model therefore predicts a rotation curve of fixed shape. This rigidity is an attractive feature, but it also means that systematic departures of an observed rotation curve from the predicted shape cannot be absorbed by the model. Such departures are already visible in the sample of ref.~\cite{harko2011bose}: the condensate profile does not describe the inner region of Holmberg~II, it follows Holmberg~I only inside about $1.5$~kpc, it matches NGC~2366 only over a limited range of radii, and for DDO~39 and DDO~154 the agreement is confined to the middle and outer parts of the curves. In ref.~\cite{harko2011bose} these discrepancies are attributed to non-circular motions, external perturbations, and the neglected baryons.

Statistical mechanics offers several generalizations of the Bose-Einstein and Fermi-Dirac statistics, including fractional and anyonic statistics, $q$-deformed bosons and fermions, and other generalized oscillator algebras \cite{wilczek1991anyons,wilczek1992disassembling,mirza2010thermodynamic,lavagno2002generalized,tuszynski1993statistical,daoud1995statistical,mohammadzadeh2017thermodynamic,chung2014q,algin2013low,stern2008anyons,lee1989anyon,manna2020anyons,tevosyan1997virial}. The $q$-deformed boson statistics generalizes the ordinary bosonic one and recovers it in the limit $q\to1$, which makes it a natural framework for extending the BEC description without abandoning its bosonic origin \cite{maleki2023beyond,maleki2025deformed,maleki2023q,maleki2024nonextensive}. Nonextensive (Tsallis) statistics \cite{salinas1999special,thurner2005nonextensive} has likewise been applied to dark energy and dark matter \cite{barboza2015dark,beck2004nonextensive,leubner2005nonextensive,zadeh2018note} and to the Gross-Pitaevskii equation \cite{maleki2023nonextensive}. Condensates obeying generalized statistics have been proposed as models of dark matter in refs.~\cite{ebadi2013infinite,gavrilik2018condensate,maleki2019deformed,maleki2022description,maleki2022mass}, and the $q$-deformed Bose statistics has recently been incorporated into the Gross-Pitaevskii framework \cite{maleki2023q}.

In this paper we ask what a condensate of $q$-deformed bosons implies for the rotation curves of dwarf galaxies. Our analysis has three parts. First, we show that in the TF limit the deformation affects the halo only through the chemical potential of the condensate. A uniform deformation parameter therefore gives exactly the profile of ref.~\cite{harko2011bose}; the $q$-deformation becomes visible in the rotation curve only if $q$ varies across the halo. Second, we adopt this local-deformation picture as a phenomenological extension of the BEC model. The density then satisfies a modified equation in which a function $\chi(r)\propto-\ln q(r)$ is added to the density inside the Laplacian. We invert this equation analytically, which gives $\chi$ in terms of the rotation velocity and its first derivative, and we use the inversion to reconstruct the deformation profile of the eight dwarf galaxies of ref.~\cite{harko2011bose}. Third, guided by the reconstruction, we parametrize $\chi(r)$ with two parameters, solve the structure equation, and fit the resulting rotation curves to the data, comparing the quality of the fits with that of the standard condensate, the NFW profile, and the cored Burkert profile.

The logical status of the first two steps should be kept in mind. The function $\chi(r)$ obtained by inversion is not a fit parameter but a free function, and the inversion yields, for any rotation curve, the $\chi(r)$ that reproduces it exactly. The agreement with the data at that stage is guaranteed and carries no evidential weight. What the reconstruction provides is a pointwise measure of the departure of each galaxy from the standard TF halo, for which $\chi$ is constant, and the empirical shape of the deformation profile. Only the third step, in which the deformation is described by a small number of parameters and compared with alternatives by a criterion that penalizes additional parameters, is a test of the model.

The paper is organized as follows. In section~\ref{2} we review the $q$-oscillator algebra and the associated $q$-calculus. In section~\ref{3} we derive the structure equation of the $q$-deformed halo in the TF approximation, its inversion in terms of the rotation curve, and its solution for a parametrized deformation profile. In section~\ref{sec:results} we apply these results to the eight dwarf galaxies and discuss their limitations. Section~\ref{4} contains our conclusions.

\section{\boldmath $q$-oscillator algebra and $q$-calculus}
\label{2}

In $q$-deformed statistics the commutation relations of the creation and annihilation operators are modified by a deformation parameter $q$, which we take to be real and positive. The symmetric $q$-boson oscillator algebra is defined in terms of the creation operator $c^\dag$, the annihilation operator $c$, and the number operator $N$ by \cite{tuszynski1993statistical,biedenharn1989quantum,macfarlane1989q}
\begin{equation}
cc^{\dag}-qc^{\dag}c=q^{-N},\qquad [c,c]=[c^{\dag},c^{\dag}]=0,
\label{number operator}
\end{equation}
\begin{equation}
[N,c^{\dag}]=c^{\dag},\qquad [N,c]=-c,
\end{equation}
\begin{equation}
c^{\dag}c=[N],\qquad cc^{\dag}=[1+N],
\label{creation operator}
\end{equation}
where the $q$-basic number is
\begin{equation}
[x]=\frac{q^{x}-q^{-x}}{q-q^{-1}}.
\label{basic numbers}
\end{equation}
The algebra is invariant under $q\to q^{-1}$ and reduces to the ordinary bosonic algebra for $q\to1$. The derivative appropriate to this algebra is the (symmetric) Jackson derivative,
\begin{equation}
{\cal{D}}^{(q)}_{x}f(x)=\frac{f(q x)-f(q^{-1}x)}{(q-q^{-1})x},
\label{the Jackson Derivative}
\end{equation}
which reduces to the ordinary derivative for $q\to1$.

The Hamiltonian of noninteracting $q$-deformed oscillators is
\begin{equation}
H=\sum_{i}(E_{i}-\mu)N_{i},
\label{Hamiltonian}
\end{equation}
where $i$ labels the single-particle states, $E_i$ is the energy of state $i$, $N_i$ is the corresponding number operator, and $\mu$ is the chemical potential. Because $H$ is linear in the number operators, the partition function has the ordinary bosonic form,
\begin{equation}
\mathrm{Tr}\,(e^{-\beta H})=\prod_i \frac{1}{1-e^{-\beta(E_i-\mu)}},
\label{p}
\end{equation}
with $\beta=1/(k_BT)$, and the deformation appears in the thermal average of $c_i^\dag c_i=[N_i]$,
\begin{equation}
[n_i]=\frac{\mathrm{Tr}\,(e^{-\beta H}c^{\dag}_{i}c_{i})}{\mathrm{Tr}\,(e^{-\beta H})}.
\label{ occupation number}
\end{equation}
The mean occupation number that follows from it is \cite{mirza2010thermodynamic,mirza2011thermodynamic}
\begin{equation}
n_{i}=\frac{1}{q-q^{-1}}\ln\left(\frac{z^{-1}e^{\beta E_{i}}-q^{-1}}{z^{-1}e^{\beta E_{i}}-q}\right),
\label{the mean value of the occupation number}
\end{equation}
where $z=e^{\beta\mu}$ is the fugacity. In contrast to ordinary bosons, for which the occupation of the ground state diverges at $z=1$, the distribution~\eqref{the mean value of the occupation number} becomes singular at a $q$-dependent value of the fugacity ($z=q^{2}$ for $q<1$ \cite{mirza2011thermodynamic}). The chemical potential of the condensed phase is therefore a function of the deformation parameter; this is the property on which the following section rests.

\section{\boldmath Thomas-Fermi halo of a $q$-deformed condensate}
\label{3}

\subsection{Structure equation}

The $q$-deformed generalization of the Gross-Pitaevskii equation is \cite{maleki2023q}
\begin{equation}
-\frac{\hbar ^{2}}{2m}\big({\cal{D}}^{(q)}\big)^{2}\psi(\vec{r}) +mV(\vec{r})\psi(\vec{r}) +U_{0}\left|\psi(\vec{r})\right|^{2}\psi(\vec{r}) =\mu\,\psi(\vec{r}),
\label{f1}
\end{equation}
where $\psi$ is the condensate wave function, normalized so that $\rho=m|\psi|^2$ is the mass density, $V$ is the gravitational potential, $m$ is the particle mass, $U_0=4\pi\hbar^{2}a/m$ is the strength of the contact interaction, and $a$ is the scattering length. The first term is the ($q$-deformed) kinetic energy, the second is the gravitational energy, and the third is the self-interaction energy.

For a halo containing a very large number of particles the kinetic (quantum pressure) term is negligible compared with the interaction term, and eq.~\eqref{f1} reduces to the Thomas-Fermi relation
\begin{equation}
\rho(\vec{r}) =\frac{m}{U_{0}} \left[ \mu-mV(\vec{r}) \right].
\label{f2}
\end{equation}
Two remarks are in order. First, the Jackson derivative has dropped out together with the kinetic term, so that in the TF limit the deformation can enter the structure of the halo only through $\mu$. Second, for the $q$-boson condensate the chemical potential is fixed by the critical value of the fugacity, $z=q^{2}$ for $q<1$ \cite{mirza2011thermodynamic,maleki2023q}, so that
\begin{equation}
\mu=k_{B}T\ln q^{2},\qquad q<1.
\label{mu}
\end{equation}

Applying the Laplacian to eq.~\eqref{f2} and using the Poisson equation $\nabla^{2}V=4\pi G\rho$ gives
\begin{equation}
\nabla ^{2}\rho +k^{2}\rho =\frac{m}{U_0}\nabla^{2}\mu ,\qquad k^{2}=\frac{4\pi G m^{2}}{U_0}=\frac{Gm^{3}}{\hbar^{2}a}.
\label{f5}
\end{equation}

\subsection{Uniform deformation}

If $q$, and therefore $\mu$, is the same everywhere in the halo, the right-hand side of eq.~\eqref{f5} vanishes and the density obeys $\nabla^{2}\rho+k^{2}\rho=0$. The regular spherically symmetric solution is
\begin{equation}
\rho(r)=\rho_c\,\frac{\sin kr}{kr},
\label{harko}
\end{equation}
with halo radius $R_h=\pi/k$, which is precisely the profile of the undeformed condensate \cite{harko2011bose}. The value of $q$ shifts the constant $\mu$, but $\rho_c$ is in any case a free parameter fixed by the total mass. We conclude that, in the TF approximation, a uniformly $q$-deformed condensate is indistinguishable from an ordinary one as far as the density profile and the rotation curve are concerned. A dependence of the profile on a constant $q$ can arise only beyond the TF approximation, through the Jackson-derivative kinetic term of eq.~\eqref{f1}, or at finite temperature through the equation of state of the non-condensed $q$-boson gas.

\subsection{Local deformation}

We now consider, as a phenomenological extension, a deformation parameter that varies slowly across the halo, $q\to q(\vec{r})$, and we continue to use eq.~\eqref{f2} locally with $\mu(\vec r)=k_BT\ln q^{2}(\vec r)$. This is an assumption, not a consequence of eq.~\eqref{f1}: a stationary solution of the GPE has a single, constant $\mu$, and a position-dependent $\mu$ is equivalent to adding an external potential to the TF relation. Its physical origin could be, for instance, a dependence of the effective statistics on the local density or temperature of the halo, but we do not attempt to model it here. Equation~\eqref{f5} becomes
\begin{equation}
\nabla ^{2}\rho +k^{2}\rho =\frac{1}{c}\,\nabla^{2}\ln q(\vec{r}),\qquad \frac1c=\frac{2mk_{B}T}{U_0}=\frac{m^{2}k_{B}T}{2\pi\hbar^{2}a}.
\label{f3}
\end{equation}
The constant $1/c$ has the dimension of a mass density. It is therefore natural to define
\begin{equation}
\chi(\vec{r})\equiv-\frac{1}{c}\ln q(\vec{r})=-\frac{m}{U_0}\,\mu(\vec r),
\label{f8}
\end{equation}
which is a mass density, is non-negative for $q\le1$, and vanishes in the Bose limit $q\to1$. A larger $\chi$ means a stronger deformation. No expansion in $1-q$ is needed; conversely $q=e^{-c\chi}$, and $q\simeq1-c\chi$ when $c\chi\ll1$. In terms of $\chi$, eq.~\eqref{f3} reads
\begin{equation}
\nabla ^{2}(\rho+\chi)+k^{2}\rho=0.
\label{f11r}
\end{equation}
Comparing eq.~\eqref{f11r} with $\nabla^{2}\rho+k^{2}\rho=0$ shows that the local deformation adds $\chi$ to the density inside the Laplacian only. Equivalently, from eqs.~\eqref{f2} and \eqref{f8}, $\rho+\chi=-(m^{2}/U_0)V$: the combination $\rho+\chi$, rather than $\rho$ itself, follows the gravitational potential.

Only $\nabla^{2}\chi$ enters eq.~\eqref{f11r}, so that $\chi$ is defined up to a harmonic function; for a regular spherically symmetric halo, up to an additive constant. This restates the result of the previous subsection: a uniform shift of $\chi$, i.e., a uniform rescaling of $q$, is unobservable in the rotation curve, and only the \emph{variation} of the deformation across the halo is physically meaningful here.

It is convenient to use the dimensionless radius $R=kr$, in terms of which the halo of eq.~\eqref{harko} extends to $R=\pi$. Since $\nabla_r^{2}=k^{2}\nabla_R^{2}$, eq.~\eqref{f11r} becomes
\begin{equation}
\nabla ^{2}_{R}(\rho+\chi)+\rho=0.
\label{f11}
\end{equation}

\subsection{Inversion in terms of the rotation curve}

Let $m(r)=4\pi\int_0^r\rho(r')r'^{2}dr'$ be the mass enclosed within radius $r$ (not to be confused with the particle mass). Then
\begin{equation}
\rho=\frac{1}{4\pi r^{2}}\frac{dm}{dr}=\frac{1}{4\pi R^{2}}\frac{dM}{dR},\qquad M(R)\equiv k^{3}m(r),
\label{f12}
\end{equation}
where $M$ has the dimension of a mass density. With $\nabla^{2}_R=R^{-2}\,d/dR\,(R^{2}\,d/dR)$, eq.~\eqref{f11} multiplied by $R^{2}$ is
\begin{equation}
\frac{d}{dR}\Big[R^{2}\frac{d}{dR}(\chi+\rho)\Big]+\frac{1}{4\pi}\frac{dM}{dR}=0.
\label{f14}
\end{equation}
A first integration gives $R^{2}\,d(\chi+\rho)/dR+M/(4\pi)=A_0$. For a halo that is regular at the centre, $M(0)=0$ and the gradient of $\chi+\rho$ is finite, hence $A_0=0$. A second integration then gives
\begin{equation}
\chi(R)+\rho(R)+\frac{1}{4\pi}\int_{0}^{R}\frac{M(R')}{R'^{2}}\,dR'=B,\qquad B=\chi(0)+\rho_c,
\label{f16}
\end{equation}
where $B$ is the additive constant discussed above.

The tangential velocity of a test particle on a circular orbit is $V^{2}(r)=Gm(r)/r$, so that
\begin{equation}
M(R)=R\,\tilde{V}^{2}(R),\qquad \tilde{V}(R)\equiv\frac{k}{G^{1/2}}\,V(R),
\label{f18}
\end{equation}
where $\tilde V^{2}$ also has the dimension of a mass density. Substituting eq.~\eqref{f18} into eqs.~\eqref{f12} and \eqref{f16} we obtain
\begin{equation}
\rho(R)=\frac{1}{4\pi}\left[\frac{\tilde{V}^{2}}{R^{2}}+\frac{2\tilde{V}}{R}\frac{d\tilde{V}}{dR}\right],
\label{rhoV}
\end{equation}
\begin{equation}
\chi(R)=B-\frac{1}{4\pi}\frac{\tilde{V}^{2}}{R^{2}}-\frac{1}{2\pi}\frac{\tilde{V}}{R}\frac{d\tilde{V}}{dR}-\frac{1}{4\pi}\int_{0}^{R}\frac{\tilde{V}^{2}(R')}{R'}\,dR'.
\label{f20}
\end{equation}
Equation~\eqref{f20} expresses the deformation profile directly in terms of the rotation curve and its first derivative. In physical units it reads $\chi(r)=B-\rho_{\rm obs}(r)-(k^{2}/4\pi G)\int_0^r V^{2}(r')\,dr'/r'$, with $\rho_{\rm obs}=(V^{2}/r^{2}+2VV'/r)/4\pi G$ the density implied by the rotation curve. As a check, for the undeformed halo~\eqref{harko} one has $M=4\pi\rho_c(\sin R-R\cos R)$ and $\int_0^R M/R'^{2}\,dR'=4\pi\rho_c\,(1-\sin R/R)$, so that eq.~\eqref{f16} gives $\chi=B-\rho_c=\chi(0)$, a constant, as it must. A rotation curve is thus compatible with the standard TF condensate if and only if the right-hand side of eq.~\eqref{f20} is independent of $R$, and the variation of $\chi$ with $R$ measures, point by point, the departure of a galaxy from that model.

Once $\chi$ is known, the deformation parameter follows from eq.~\eqref{f8}. Using $\hbar^{2}a=Gm^{3}/k^{2}$ to eliminate the scattering length, $c=2\pi Gm/(k^{2}k_BT)$ and
\begin{equation}
\frac{q(R)}{q(R_0)}=\exp\!\left\{-\frac{2\pi G m}{k^{2}k_{B}T}\,\big[\chi(R)-\chi(R_0)\big]\right\},
\label{qprofile}
\end{equation}
for any reference radius $R_0$. Because of the freedom in $B$, the rotation curve determines only ratios of $q$ at different radii, and converting $\chi$ into $q$ requires, in addition to $k$, the ratio $m/T$, which the rotation curve does not constrain.

\subsection{A parametrized deformation profile}\label{sec:model}

Equation~\eqref{f20} is an inversion, not a model: it returns a $\chi(R)$ for any rotation curve. To obtain a model that can be fitted and compared with others, $\chi(R)$ must be specified by a small number of parameters and eq.~\eqref{f11} solved for the density. Writing $w(R)\equiv R\,[\rho(R)+\chi(R)]$, eq.~\eqref{f11} becomes $w''+w=R\chi$, and the solution regular at the centre with central density $\rho_c$ is
\begin{equation}
w(R)=\big[\rho_c+\chi(0)\big]\sin R+\int_0^R\sin(R-t)\,t\,\chi(t)\,dt,
\qquad \rho=\frac{w}{R}-\chi,\qquad M=4\pi\,(w-Rw').
\label{green}
\end{equation}
The halo ends at the first zero of $\rho$, beyond which the rotation curve is Keplerian. For constant $\chi$ eq.~\eqref{green} reduces to eq.~\eqref{harko}. Anticipating the result of the reconstruction in section~\ref{sec:results}, where $\chi$ is found to be depressed in the central region and approximately constant outside it, we adopt the simplest smooth profile of this kind,
\begin{equation}
\chi(R)=\chi_\infty-A\,\exp\!\left(-\frac{R^{2}}{2s^{2}}\right),\qquad A\ge0,
\label{gauss}
\end{equation}
where $A$ is the depth of the depression, $s$ is its width in units of $1/k$ (so that $\sigma=s/k$ is the width in kpc), and the constant $\chi_\infty$ drops out of the rotation curve. The resulting model has four parameters, $\rho_c$, $R_h=\pi/k$, $A$, and $s$, and reduces to the standard condensate for $A=0$. By eq.~\eqref{qprofile} it corresponds to $q(R)/q_\infty=\exp[cA\,e^{-R^{2}/2s^{2}}]$: the deformation is uniform in the envelope and weaker, i.e., $q$ closer to unity, in the core.

\section{Application to eight dwarf galaxies}\label{sec:results}

\subsection{Data}

We use the rotation curves of the eight dwarf galaxies considered in ref.~\cite{harko2011bose}: IC~2574, NGC~2366, Holmberg~I (Ho~I), Holmberg~II (Ho~II), DDO~53, and M81~dwB from the THINGS dwarf sample \cite{oh2011dark}, DDO~39 from ref.~\cite{swaters2003central}, and DDO~154 from ref.~\cite{carignan1989optical}. The data points were read from Fig.~1 of ref.~\cite{harko2011bose}, which is stored as vector graphics, so that the centres of the plotted symbols can be recovered without the uncertainty of a manual digitization. The marker placed at the origin of each panel was discarded, which leaves $193$ points in total (table~\ref{tab:fits}). As a check of the axis calibration, the solid curves of the same figure were extracted in the same way and compared with the TF rotation curve computed from the parameters of table~\ref{tab:halo}; they agree to better than $0.2$~km\,s$^{-1}$ for six galaxies and to $2$~km\,s$^{-1}$ for Ho~II and DDO~39, whose plotted curves appear to have been drawn with slightly different parameters. The figure carries no error bars, and, as in ref.~\cite{harko2011bose}, the contribution of baryonic matter to the rotation curves is neglected.

\begin{table}[tbp]
\centering
\begin{tabular}{lcccc}
\hline
Galaxy & $R_h$ (kpc) & $\rho_c$ ($10^{-24}$ g\,cm$^{-3}$) & $k$ (kpc$^{-1}$) & Data \\
\hline
IC~2574 & 12.6 & 0.44 & 0.249 & \cite{oh2011dark} \\
NGC~2366 & 6.5 & 0.89 & 0.483 & \cite{oh2011dark} \\
Ho~I & 2.2 & 3.7 & 1.428 & \cite{oh2011dark} \\
Ho~II & 7.0 & 0.35 & 0.449 & \cite{oh2011dark} \\
DDO~53 & 2.54 & 1.8 & 1.237 & \cite{oh2011dark} \\
M81~dwB & 1.0 & 18.0 & 3.142 & \cite{oh2011dark} \\
DDO~39 & 9.5 & 1.04 & 0.331 & \cite{swaters2003central} \\
DDO~154 & 7.5 & 0.60 & 0.419 & \cite{carignan1989optical} \\
\hline
\end{tabular}
\caption{\label{tab:halo}Halo parameters of the standard Thomas-Fermi condensate given in ref.~\cite{harko2011bose} (chosen there by inspection, not fitted): halo radius $R_h$ and central density $\rho_c$. The third column gives $k=\pi/R_h$, used to define $R=kr$ in the reconstruction of section~\ref{sec:recon}, and the last column the source of the rotation curve.}
\end{table}

\subsection{Reconstructed deformation profiles}\label{sec:recon}

\begin{figure}[tbp]
\centering
\includegraphics[width=.9\textwidth]{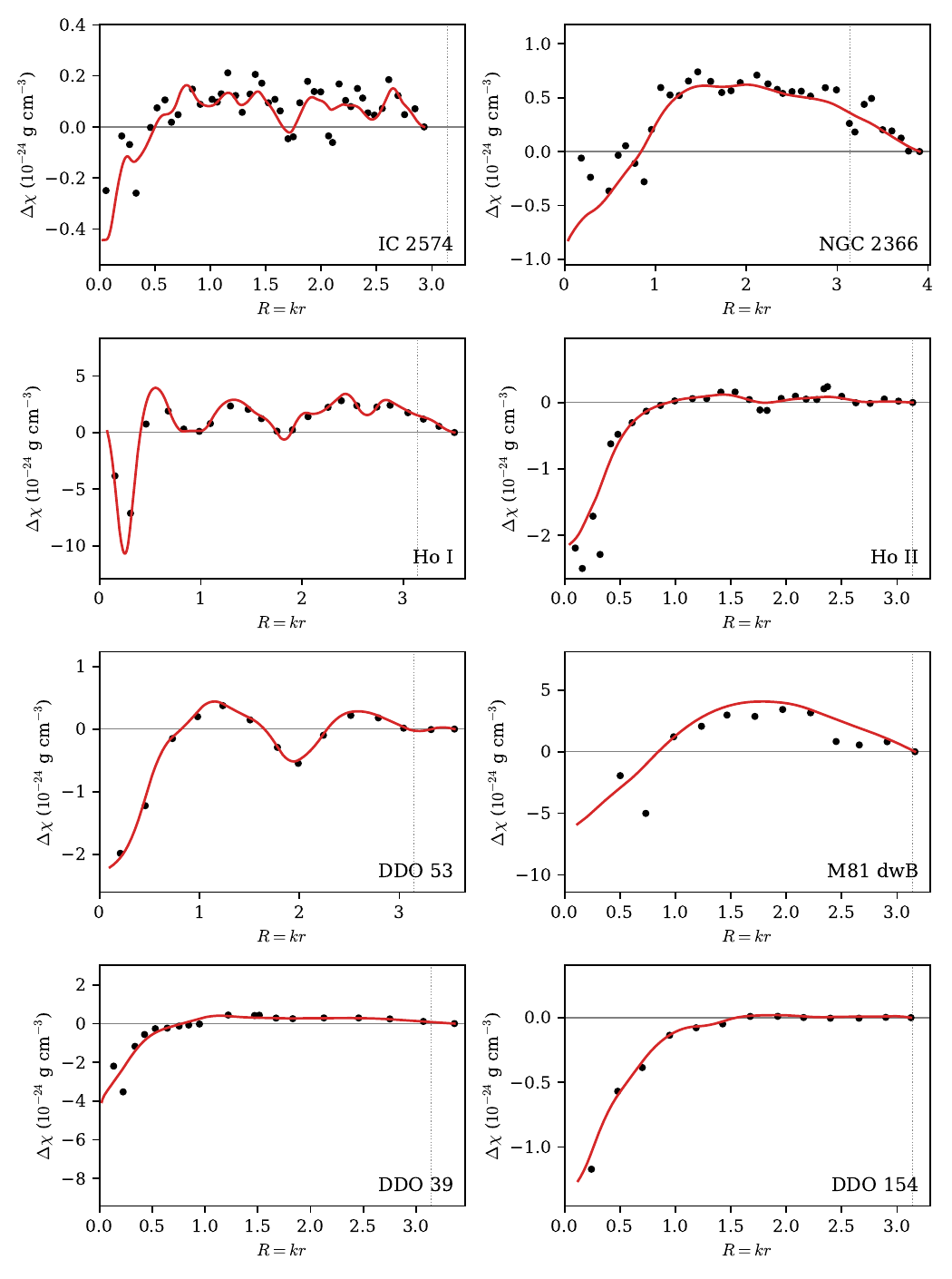}
\caption{\label{fig:chi}Deformation profile $\Delta\chi(R)=\chi(R)-\chi(R_{\rm out})$ reconstructed from the rotation curves with eq.~\eqref{f20}, as a function of the dimensionless radius $R=kr=\pi r/R_h$, with $R_h$ from table~\ref{tab:halo}. Points: finite-difference estimate at the measured radii. Curves: estimate from a smoothing spline of the rotation curve. The dotted vertical line marks the edge $R=\pi$ of the standard condensate halo. For the standard Thomas-Fermi condensate $\Delta\chi$ would vanish identically.}
\end{figure}

For each galaxy we evaluate eq.~\eqref{f20} with $k=\pi/R_h$ and the halo radius of table~\ref{tab:halo}. Since $\chi$ is defined up to a constant, we show $\Delta\chi(R)\equiv\chi(R)-\chi(R_{\rm out})$, where $R_{\rm out}$ is the outermost measured point. The derivative and the integral in eq.~\eqref{f20} are computed in two ways: directly from the data points, by finite differences and the trapezoidal rule, and from a smoothing spline of $V(r)$ whose smoothing parameter is chosen by generalized cross-validation. The results are shown in figure~\ref{fig:chi}.

The profiles are not constant, and they depart from a constant in the same way in every galaxy: $\Delta\chi$ is \emph{negative} in the inner region, $R\lesssim1$, and rises towards the outer value. In Ho~II, DDO~39, DDO~154, and, within a larger scatter, IC~2574 and DDO~53, the profile is flat for $R\gtrsim1$; in NGC~2366 and M81~dwB it passes through a maximum and declines again towards the edge; in Ho~I, whose outer rotation curve is strongly disturbed \cite{harko2011bose}, it oscillates. The origin of the central depression is simple. With $\chi=B-\rho_{\rm obs}-(k^{2}/4\pi G)\int V^{2}dr/r$, a negative $\Delta\chi$ at small radii means that the density implied by the rotation curve rises towards the centre more steeply than the $\sin R/R$ core allows; this is the same feature that makes the inner parts of the observed curves lie above the TF prediction in Fig.~1 of ref.~\cite{harko2011bose}. In the language of the model, the chemical potential $\mu=-(U_0/m)\chi$ must be higher in the core than in the envelope, and by eq.~\eqref{qprofile} $q$ is larger there: the deformation is weaker in the core and approximately uniform outside it. The sign of the effect does not depend on the adopted halo radius: repeating the reconstruction with the fitted radii of table~\ref{tab:pars} instead of those of table~\ref{tab:halo} changes the depth of the depression by about $30\%$ at most.

The innermost points must be read with caution. There $\rho_{\rm obs}\propto V^{2}/r^{2}$ amplifies both the noise of the data and any systematic error of the inner rotation curve, and it is at small radii that non-circular motions and the neglected baryons matter most.

\begin{figure}[tbp]
\centering
\includegraphics[width=.85\textwidth]{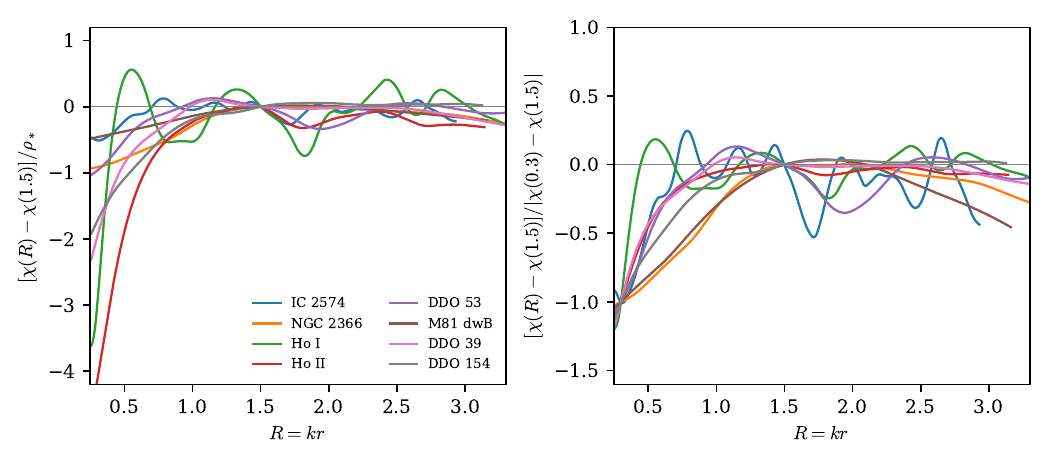}
\caption{\label{fig:collapse}Test of universality of the deformation profile (smoothing-spline estimates of figure~\ref{fig:chi}, referred to their value at $R=1.5$). Left: in units of $\rho_*=k^{2}V_{\max}^{2}/4\pi G$. Right: normalized to the depth of each profile between $R=0.3$ and $R=1.5$.}
\end{figure}

To test whether the profiles are universal, we compare them in dimensionless form in figure~\ref{fig:collapse}. The natural density scale that can be formed from the data and $k$ is $\rho_*\equiv k^{2}V_{\max}^{2}/4\pi G$, which agrees with the central densities of table~\ref{tab:halo} to within $45\%$. In units of $\rho_*$ the profiles do not collapse: the depth of the depression, measured between $R=0.3$ and $R=1.5$, ranges from about $0.5\rho_*$ (M81~dwB, IC~2574) to $4\rho_*$ (Ho~II). When each profile is instead normalized to its own depth, the curves are similar for $R\lesssim1.5$, but this normalization removes one degree of freedom per galaxy and the agreement is correspondingly less informative. We conclude that the reconstructed profiles share a common qualitative shape, a central depression of width $\Delta R\sim1$ followed by a plateau, but that neither its amplitude nor, as shown below, its width is universal.

\subsection{Fits of the rotation curves}

\begin{figure}[tbp]
\centering
\includegraphics[width=.9\textwidth]{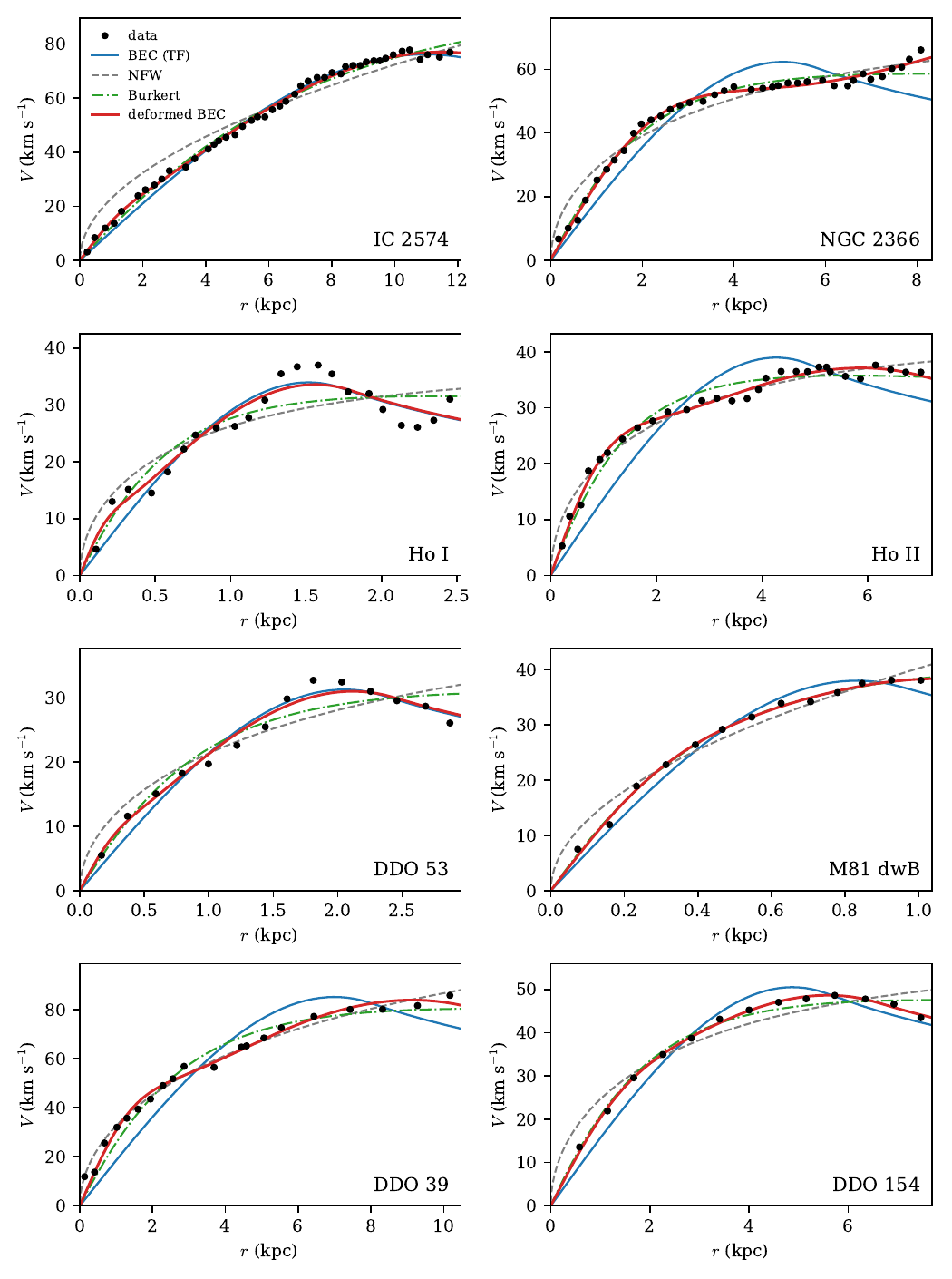}
\caption{\label{fig:fits}Rotation curves of the eight dwarf galaxies (points, from Fig.~1 of ref.~\cite{harko2011bose}) and least-squares fits with the standard Thomas-Fermi condensate (blue solid), the NFW profile (grey dashed), the Burkert profile (green dash-dotted), and the locally deformed condensate with the Gaussian deformation profile of eq.~\eqref{gauss} (red solid).}
\end{figure}

We now fit the rotation curves with four models: the standard TF condensate, eq.~\eqref{harko}, with free $\rho_c$ and $R_h$; the NFW profile, with free $\rho_s$ and $r_s$; the cored Burkert profile, $\rho=\rho_0/[(1+r/r_0)(1+r^{2}/r_0^{2})]$, with free $\rho_0$ and $r_0$; and the locally deformed condensate of section~\ref{sec:model}, with free $\rho_c$, $R_h$, $A$, and $s$. Because no error bars are available, the fits are unweighted least-squares fits, and we compare the models through the rms residual and the Bayesian information criterion for Gaussian errors of unknown common variance, ${\rm BIC}=N\ln({\rm RSS}/N)+p\ln N$, where RSS is the residual sum of squares, $N$ the number of points, and $p$ the number of parameters. A difference of BIC larger than about $10$ is conventionally regarded as strong evidence. The results are given in tables~\ref{tab:fits} and \ref{tab:pars} and in figure~\ref{fig:fits}.

\begin{table}[tbp]
\centering
\small\setlength{\tabcolsep}{4pt}
\begin{tabular}{lcccccccccc}
\hline
 & & \multicolumn{5}{c}{rms residual} & \multicolumn{4}{c}{BIC} \\
Galaxy & $N$ & BEC \cite{harko2011bose} & BEC & NFW & Burkert & Deformed & BEC & NFW & Burkert & Deformed \\
\hline
IC~2574 & 45 & 2.13 & 1.90 & 4.91 & 1.87 & 1.15 & 65.5 & 150.8 & 63.7 & \textbf{27.6} \\
NGC~2366 & 37 & 7.69 & 5.82 & 3.61 & 2.09 & 1.24 & 137.5 & 102.2 & 61.9 & \textbf{30.5} \\
Ho~I & 22 & 5.64 & 2.51 & 4.16 & 3.45 & 2.18 & 46.7 & 68.9 & 60.7 & \textbf{46.7} \\
Ho~II & 29 & 6.15 & 4.33 & 1.61 & 1.61 & 0.92 & 91.7 & 34.3 & 34.5 & \textbf{8.7} \\
DDO~53 & 14 & 1.88 & 1.59 & 3.09 & 2.33 & 1.26 & 18.2 & 36.9 & 29.0 & \textbf{17.0} \\
M81~dwB & 13 & 1.96 & 1.74 & 1.89 & 0.60 & 0.60 & 19.5 & 21.7 & \textbf{-8.0} & -2.9 \\
DDO~39 & 20 & 9.48 & 8.45 & 2.45 & 4.01 & 2.56 & 91.4 & \textbf{41.8} & 61.5 & 49.5 \\
DDO~154 & 13 & 6.06 & 2.73 & 3.22 & 1.53 & 0.55 & 31.2 & 35.5 & 16.1 & \textbf{-5.2} \\
\hline
Total & 193 & & & & & & 501.8 & 492.0 & 319.5 & \textbf{171.9} \\
\hline
\end{tabular}
\caption{\label{tab:fits}Quality of the fits. $N$ is the number of data points. The rms residuals are in km\,s$^{-1}$; the column ``BEC \cite{harko2011bose}'' refers to the unfitted curves with the parameters of table~\ref{tab:halo}. The lowest BIC of each galaxy is in bold. The number of fitted parameters is 2 for BEC, NFW, and Burkert, and 4 for the deformed condensate.}
\end{table}

\begin{table}[tbp]
\centering
\begin{tabular}{lccccccccc}
\hline
 & \multicolumn{2}{c}{Standard BEC} & \multicolumn{7}{c}{Deformed BEC} \\
Galaxy & $\rho_c$ & $R_h$ & $\rho_c$ & $R_h$ & $A$ & $A/\rho_c$ & $s$ & $\sigma$ & $r_{\rm edge}$ \\
\hline
IC~2574 & 0.43 & 12.58 & 0.91 & 13.3 & 0.50 & 0.54 & 0.23 & 0.97 & 13.2 \\
NGC~2366 & 1.34 & 5.81 & 2.41 & $\gtrsim20$ & 2.23 & 0.93 & 0.10 & 1.61 & $\gtrsim20$ \\
Ho~I & 4.45 & 1.74 & 16.77 & 1.8 & 12.26 & 0.73 & 0.22 & 0.13 & 1.8 \\
Ho~II & 0.74 & 4.88 & 2.72 & 7.7 & 2.27 & 0.84 & 0.30 & 0.74 & 7.0 \\
DDO~53 & 2.06 & 2.35 & 5.44 & 2.5 & 3.40 & 0.62 & 0.25 & 0.20 & 2.4 \\
M81~dwB & 18.44 & 0.95 & 28.21 & 1.4 & 16.63 & 0.59 & 0.51 & 0.23 & 1.3 \\
DDO~39 & 1.32 & 8.01 & 4.48 & 11.2 & 3.58 & 0.80 & 0.28 & 1.01 & 10.7 \\
DDO~154 & 0.95 & 5.58 & 1.86 & 7.1 & 1.15 & 0.62 & 0.45 & 1.03 & 6.6 \\
\hline
\end{tabular}
\caption{\label{tab:pars}Best-fit parameters of the standard and of the locally deformed condensate. Densities are in $10^{-24}$ g\,cm$^{-3}$ and lengths in kpc. $\sigma=sR_h/\pi$ is the width of the central depression of $\chi$ in physical units and $r_{\rm edge}$ the radius at which the density of the deformed halo vanishes. For NGC~2366 the halo radius of the deformed model is not bounded from above by the data; the quoted $s$ refers to the formal best fit, $R_h=48.7$~kpc, and only $\sigma$ is well determined.}
\end{table}

The standard condensate, even with both parameters fitted, leaves rms residuals between $1.6$ and $8.5$~km\,s$^{-1}$, and in NGC~2366, Ho~II, DDO~39, and DDO~154 it misses the shape of the curve systematically, rising too slowly at small radii and turning over too early. The deformed condensate reduces the residuals to $0.6$--$2.6$~km\,s$^{-1}$ and removes this systematic pattern. The improvement is paid for with two additional parameters, which the BIC takes into account. According to it, the deformed condensate is strongly preferred over all three two-parameter profiles in four galaxies (IC~2574, NGC~2366, Ho~II, and DDO~154, with $\Delta{\rm BIC}$ between $21$ and $36$ relative to the best of them); it is statistically indistinguishable from the standard condensate in Ho~I and DDO~53 ($|\Delta{\rm BIC}|\le1.2$); and it is disfavoured in two galaxies, relative to the Burkert profile in M81~dwB ($\Delta{\rm BIC}=5$) and to the NFW profile in DDO~39 ($\Delta{\rm BIC}=8$). Summed over the sample, the BIC is $172$ for the deformed condensate, against $320$ for Burkert, $492$ for NFW, and $502$ for the standard condensate. Among the two-parameter profiles no single one is best throughout: the Burkert profile has the lowest BIC in four galaxies, the standard condensate in two, the NFW profile in one, and in Ho~II the NFW and Burkert profiles are tied.

The fitted parameters show what the deformation does. The depth $A$ of the depression of $\chi$ is between $54\%$ and $93\%$ of the central density, so the deformation is not a small correction: the fitted halo consists of a compact central concentration, of width $\sigma$ between $0.13$ and $1.6$~kpc, embedded in an extended TF halo. The outer radii of the deformed haloes are close to the values of table~\ref{tab:halo} except for NGC~2366, whose rotation curve is still rising at the last measured point and constrains only $R_h\gtrsim20$~kpc. In Ho~I, DDO~53, and DDO~154 the edge of the fitted halo lies inside the last measured radius, and the declining outer part of the curve is described by the Keplerian fall-off.

The width of the depression is not universal. In units of $1/k$ it ranges from $s=0.22$ to $s=0.51$ (leaving aside NGC~2366, where $k$ is not determined), and in physical units by more than a factor of ten. Forcing a common value of $s$ on all galaxies gives $s=0.30$ and a total BIC of $203$ (25 parameters), compared with $172$ when $s$ is free (32 parameters), so that the data reject a universal width. Together with the absence of a collapse in figure~\ref{fig:collapse}, this means that in its present form the model requires two additional parameters per galaxy.

\subsection{Limitations}

Several limitations qualify these results. (i) The data were read from a published figure without error bars, the fits are unweighted, and neighbouring points of a tilted-ring rotation curve are not independent; the BIC differences should therefore be taken as indicative. (ii) The Gaussian form~\eqref{gauss} was suggested by the reconstruction of section~\ref{sec:recon} and then fitted to the same rotation curves, so that the fits do not constitute an independent test of that form; and we have compared the four-parameter model only with two-parameter profiles, not with more flexible empirical ones. (iii) The local deformation $q(\vec r)$ is a phenomenological assumption whose microscopic origin remains to be established, and the fits show that the corresponding modification of the halo is of order unity in the core. A conventional explanation of the same central concentration, by the baryonic component and by non-circular motions, both neglected here and both most important at small radii \cite{harko2011bose}, has not been excluded. (iv) The chemical potential~\eqref{mu} is a finite-temperature quantity, whereas eq.~\eqref{f1} describes a pure condensate; a consistent treatment should include the thermal cloud. (v) As in the standard TF condensate, $k^{2}=Gm^{3}/\hbar^{2}a$ depends only on the properties of the dark matter particle and should be the same in all galaxies, whereas the fitted halo radii differ by an order of magnitude. This difficulty is inherited from the standard model and is not resolved by the deformation considered here.

\section{Conclusions}
We have studied a dark matter halo made of a condensate of $q$-deformed bosons in the Thomas-Fermi approximation, starting from the $q$-deformed Gross-Pitaevskii equation. Our first result is negative but, we believe, useful: in this approximation the deformation enters only through the chemical potential, and a spatially uniform $q$ reproduces exactly the density profile and the rotation curve of the ordinary condensate. A uniform $q$-deformation therefore cannot improve the description of rotation curves at the Thomas-Fermi level.

Allowing the deformation parameter to vary across the halo leads to the modified structure equation $\nabla^{2}(\rho+\chi)+k^{2}\rho=0$, with $\chi=-(1/c)\ln q$ a mass density equivalent to a position-dependent shift of the chemical potential. We have inverted this equation in closed form, eq.~\eqref{f20}, which gives $\chi$ in terms of the rotation curve and provides a pointwise diagnostic of the departure of a galaxy from the standard condensate halo. Applied to the rotation curves of eight dwarf galaxies, the inversion shows the same pattern in all of them: $\chi$ is depressed in the inner region, $kr\lesssim1$, and approximately constant farther out in most. The standard condensate thus describes the outer halo, while the observed central concentration requires a higher chemical potential, i.e., a weaker deformation, in the core.

Modelling the depression by a Gaussian, at the cost of two additional parameters, gives rotation curves that fit the data with rms residuals of $0.6$--$2.6$~km\,s$^{-1}$. By the Bayesian information criterion this locally deformed condensate is preferred over the standard condensate, the NFW profile, and the Burkert profile in four of the eight galaxies, is indistinguishable from the best of them in two, and is disfavoured in two. At the same time, the depth and the width of the depression differ from galaxy to galaxy and do not follow a universal law in the variables natural to the model, and the depth is comparable to the central density itself.

We therefore regard the locally deformed condensate as a flexible and economical description of these rotation curves, not as established evidence for $q$-deformed dark matter. Three steps would decide the matter. The first is to repeat the analysis on rotation curves with published uncertainties and with the baryonic contribution subtracted, on a larger sample that includes low-surface-brightness galaxies, and to include more flexible empirical profiles in the comparison. The second is to explain the central depression of $\chi$, either by deriving a relation between the local deformation and the local state of the halo, which would turn $A$ and $s$ from fitted into predicted quantities, or by showing that baryons and non-circular motions account for it. The third is to obtain a model with a constant $q$ by going beyond the present approximation, either by retaining the Jackson-derivative kinetic term or by solving the hydrostatic equilibrium of the halo with the finite-temperature equation of state of the $q$-boson gas, in which $q$ modifies the shape of the density profile directly.


\bibliography{refs}

\begin{thebibliography}{40}%
\makeatletter
\providecommand \@ifxundefined [1]{%
 \@ifx{#1\undefined}
}%
\providecommand \@ifnum [1]{%
 \ifnum #1\expandafter \@firstoftwo
 \else \expandafter \@secondoftwo
 \fi
}%
\providecommand \@ifx [1]{%
 \ifx #1\expandafter \@firstoftwo
 \else \expandafter \@secondoftwo
 \fi
}%
\providecommand \natexlab [1]{#1}%
\providecommand \enquote  [1]{``#1''}%
\providecommand \bibnamefont  [1]{#1}%
\providecommand \bibfnamefont [1]{#1}%
\providecommand \citenamefont [1]{#1}%
\providecommand \href@noop [0]{\@secondoftwo}%
\providecommand \href [0]{\begingroup \@sanitize@url \@href}%
\providecommand \@href[1]{\@@startlink{#1}\@@href}%
\providecommand \@@href[1]{\endgroup#1\@@endlink}%
\providecommand \@sanitize@url [0]{\catcode `\\12\catcode `\$12\catcode
  `\&12\catcode `\#12\catcode `\^12\catcode `\_12\catcode `\%12\relax}%
\providecommand \@@startlink[1]{}%
\providecommand \@@endlink[0]{}%
\providecommand \url  [0]{\begingroup\@sanitize@url \@url }%
\providecommand \@url [1]{\endgroup\@href {#1}{\urlprefix }}%
\providecommand \urlprefix  [0]{URL }%
\providecommand \Eprint [0]{\href }%
\providecommand \doibase [0]{https://doi.org/}%
\providecommand \selectlanguage [0]{\@gobble}%
\providecommand \bibinfo  [0]{\@secondoftwo}%
\providecommand \bibfield  [0]{\@secondoftwo}%
\providecommand \translation [1]{[#1]}%
\providecommand \BibitemOpen [0]{}%
\providecommand \bibitemStop [0]{}%
\providecommand \bibitemNoStop [0]{.\EOS\space}%
\providecommand \EOS [0]{\spacefactor3000\relax}%
\providecommand \BibitemShut  [1]{\csname bibitem#1\endcsname}%
\let\auto@bib@innerbib\@empty
\bibitem [{\citenamefont {Peebles}(2015)}]{peebles2015physical}%
  \BibitemOpen
  \bibfield  {author} {\bibinfo {author} {\bibfnamefont {P.~J.~E.}\
  \bibnamefont {Peebles}},\ }\href@noop {} {\emph {\bibinfo {title} {Physical
  cosmology}}}\ (\bibinfo  {publisher} {Princeton University Press},\ \bibinfo
  {year} {2015})\BibitemShut {NoStop}%
\bibitem [{\citenamefont {Weinberg}(2008)}]{weinberg2008cosmology}%
  \BibitemOpen
  \bibfield  {author} {\bibinfo {author} {\bibfnamefont {S.}~\bibnamefont
  {Weinberg}},\ }\href@noop {} {\emph {\bibinfo {title} {Cosmology}}}\
  (\bibinfo  {publisher} {OUP Oxford},\ \bibinfo {year} {2008})\BibitemShut
  {NoStop}%
\bibitem [{\citenamefont {Trimble}(1987)}]{trimble1987existence}%
  \BibitemOpen
  \bibfield  {author} {\bibinfo {author} {\bibfnamefont {V.}~\bibnamefont
  {Trimble}},\ }\bibfield  {title} {\bibinfo {title} {Existence and nature of
  dark matter in the universe},\ }\href@noop {} {\bibfield  {journal} {\bibinfo
   {journal} {Annual review of astronomy and astrophysics}\ }\textbf {\bibinfo
  {volume} {25}},\ \bibinfo {pages} {425} (\bibinfo {year} {1987})}\BibitemShut
  {NoStop}%
\bibitem [{\citenamefont {Navarro}\ \emph {et~al.}(1997)\citenamefont
  {Navarro}, \citenamefont {Frenk},\ and\ \citenamefont
  {White}}]{navarro1997universal}%
  \BibitemOpen
  \bibfield  {author} {\bibinfo {author} {\bibfnamefont {J.~F.}\ \bibnamefont
  {Navarro}}, \bibinfo {author} {\bibfnamefont {C.~S.}\ \bibnamefont {Frenk}},\
  and\ \bibinfo {author} {\bibfnamefont {S.~D.}\ \bibnamefont {White}},\
  }\bibfield  {title} {\bibinfo {title} {A universal density profile from
  hierarchical clustering},\ }\href@noop {} {\bibfield  {journal} {\bibinfo
  {journal} {The Astrophysical Journal}\ }\textbf {\bibinfo {volume} {490}},\
  \bibinfo {pages} {493} (\bibinfo {year} {1997})}\BibitemShut {NoStop}%
\bibitem [{\citenamefont {Harko}(2011)}]{harko2011bose}%
  \BibitemOpen
  \bibfield  {author} {\bibinfo {author} {\bibfnamefont {T.}~\bibnamefont
  {Harko}},\ }\bibfield  {title} {\bibinfo {title} {Bose-einstein condensation
  of dark matter solves the core/cusp problem},\ }\href@noop {} {\bibfield
  {journal} {\bibinfo  {journal} {Journal of Cosmology and Astroparticle
  Physics}\ }\textbf {\bibinfo {volume} {2011}}\bibinfo  {number} { (05)},\
  \bibinfo {pages} {022}}\BibitemShut {NoStop}%
\bibitem [{\citenamefont {Wilczek}(1991)}]{wilczek1991anyons}%
  \BibitemOpen
\bibfield  {number} {  }\bibfield  {author} {\bibinfo {author} {\bibfnamefont
  {F.}~\bibnamefont {Wilczek}},\ }\bibfield  {title} {\bibinfo {title}
  {Anyons},\ }\href@noop {} {\bibfield  {journal} {\bibinfo  {journal}
  {Scientific American}\ }\textbf {\bibinfo {volume} {264}},\ \bibinfo {pages}
  {58} (\bibinfo {year} {1991})}\BibitemShut {NoStop}%
\bibitem [{\citenamefont {Wilczek}(1992)}]{wilczek1992disassembling}%
  \BibitemOpen
  \bibfield  {author} {\bibinfo {author} {\bibfnamefont {F.}~\bibnamefont
  {Wilczek}},\ }\bibfield  {title} {\bibinfo {title} {Disassembling anyons},\
  }\href@noop {} {\bibfield  {journal} {\bibinfo  {journal} {Physical review
  letters}\ }\textbf {\bibinfo {volume} {69}},\ \bibinfo {pages} {132}
  (\bibinfo {year} {1992})}\BibitemShut {NoStop}%
\bibitem [{\citenamefont {Mirza}\ and\ \citenamefont
  {Mohammadzadeh}(2010)}]{mirza2010thermodynamic}%
  \BibitemOpen
  \bibfield  {author} {\bibinfo {author} {\bibfnamefont {B.}~\bibnamefont
  {Mirza}}\ and\ \bibinfo {author} {\bibfnamefont {H.}~\bibnamefont
  {Mohammadzadeh}},\ }\bibfield  {title} {\bibinfo {title} {Thermodynamic
  geometry of fractional statistics},\ }\href@noop {} {\bibfield  {journal}
  {\bibinfo  {journal} {Physical Review E}\ }\textbf {\bibinfo {volume} {82}},\
  \bibinfo {pages} {031137} (\bibinfo {year} {2010})}\BibitemShut {NoStop}%
\bibitem [{\citenamefont {Lavagno}\ and\ \citenamefont
  {Swamy}(2002)}]{lavagno2002generalized}%
  \BibitemOpen
  \bibfield  {author} {\bibinfo {author} {\bibfnamefont {A.}~\bibnamefont
  {Lavagno}}\ and\ \bibinfo {author} {\bibfnamefont {P.~N.}\ \bibnamefont
  {Swamy}},\ }\bibfield  {title} {\bibinfo {title} {Generalized thermodynamics
  of q-deformed bosons and fermions},\ }\href@noop {} {\bibfield  {journal}
  {\bibinfo  {journal} {Physical Review E}\ }\textbf {\bibinfo {volume} {65}},\
  \bibinfo {pages} {036101} (\bibinfo {year} {2002})}\BibitemShut {NoStop}%
\bibitem [{\citenamefont {Tuszy{\'n}ski}\ \emph {et~al.}(1993)\citenamefont
  {Tuszy{\'n}ski}, \citenamefont {Rubin}, \citenamefont {Meyer},\ and\
  \citenamefont {Kibler}}]{tuszynski1993statistical}%
  \BibitemOpen
  \bibfield  {author} {\bibinfo {author} {\bibfnamefont {J.}~\bibnamefont
  {Tuszy{\'n}ski}}, \bibinfo {author} {\bibfnamefont {J.}~\bibnamefont
  {Rubin}}, \bibinfo {author} {\bibfnamefont {J.}~\bibnamefont {Meyer}},\ and\
  \bibinfo {author} {\bibfnamefont {M.}~\bibnamefont {Kibler}},\ }\bibfield
  {title} {\bibinfo {title} {Statistical mechanics of a q-deformed boson gas},\
  }\href@noop {} {\bibfield  {journal} {\bibinfo  {journal} {Physics Letters
  A}\ }\textbf {\bibinfo {volume} {175}},\ \bibinfo {pages} {173} (\bibinfo
  {year} {1993})}\BibitemShut {NoStop}%
\bibitem [{\citenamefont {Daoud}\ and\ \citenamefont
  {Kibler}(1995)}]{daoud1995statistical}%
  \BibitemOpen
  \bibfield  {author} {\bibinfo {author} {\bibfnamefont {M.}~\bibnamefont
  {Daoud}}\ and\ \bibinfo {author} {\bibfnamefont {M.}~\bibnamefont {Kibler}},\
  }\bibfield  {title} {\bibinfo {title} {Statistical mechanics of qp-bosons in
  d dimensions},\ }\href@noop {} {\bibfield  {journal} {\bibinfo  {journal}
  {Physics Letters A}\ }\textbf {\bibinfo {volume} {206}},\ \bibinfo {pages}
  {13} (\bibinfo {year} {1995})}\BibitemShut {NoStop}%
\bibitem [{\citenamefont {Mohammadzadeh}\ \emph {et~al.}(2017)\citenamefont
  {Mohammadzadeh}, \citenamefont {Azizian-Kalandaragh}, \citenamefont
  {Cheraghpour},\ and\ \citenamefont {Adli}}]{mohammadzadeh2017thermodynamic}%
  \BibitemOpen
  \bibfield  {author} {\bibinfo {author} {\bibfnamefont {H.}~\bibnamefont
  {Mohammadzadeh}}, \bibinfo {author} {\bibfnamefont {Y.}~\bibnamefont
  {Azizian-Kalandaragh}}, \bibinfo {author} {\bibfnamefont {N.}~\bibnamefont
  {Cheraghpour}},\ and\ \bibinfo {author} {\bibfnamefont {F.}~\bibnamefont
  {Adli}},\ }\bibfield  {title} {\bibinfo {title} {Thermodynamic geometry,
  condensation and debye model of two-parameter deformed statistics},\
  }\href@noop {} {\bibfield  {journal} {\bibinfo  {journal} {Journal of
  Statistical Mechanics: Theory and Experiment}\ }\textbf {\bibinfo {volume}
  {2017}},\ \bibinfo {pages} {083104} (\bibinfo {year} {2017})}\BibitemShut
  {NoStop}%
\bibitem [{\citenamefont {Chung}\ and\ \citenamefont
  {Algin}(2014)}]{chung2014q}%
  \BibitemOpen
  \bibfield  {author} {\bibinfo {author} {\bibfnamefont {W.~S.}\ \bibnamefont
  {Chung}}\ and\ \bibinfo {author} {\bibfnamefont {A.}~\bibnamefont {Algin}},\
  }\bibfield  {title} {\bibinfo {title} {q-deformed tamm--dancoff oscillators:
  Coherent state and thermostatistics},\ }\href@noop {} {\bibfield  {journal}
  {\bibinfo  {journal} {International Journal of Modern Physics B}\ }\textbf
  {\bibinfo {volume} {28}},\ \bibinfo {pages} {1450130} (\bibinfo {year}
  {2014})}\BibitemShut {NoStop}%
\bibitem [{\citenamefont {Algin}\ and\ \citenamefont
  {Ilik}(2013)}]{algin2013low}%
  \BibitemOpen
  \bibfield  {author} {\bibinfo {author} {\bibfnamefont {A.}~\bibnamefont
  {Algin}}\ and\ \bibinfo {author} {\bibfnamefont {E.}~\bibnamefont {Ilik}},\
  }\bibfield  {title} {\bibinfo {title} {Low-temperature thermostatistics of
  tamm--dancoff deformed boson oscillators},\ }\href@noop {} {\bibfield
  {journal} {\bibinfo  {journal} {Physics Letters A}\ }\textbf {\bibinfo
  {volume} {377}},\ \bibinfo {pages} {1797} (\bibinfo {year}
  {2013})}\BibitemShut {NoStop}%
\bibitem [{\citenamefont {Stern}(2008)}]{stern2008anyons}%
  \BibitemOpen
  \bibfield  {author} {\bibinfo {author} {\bibfnamefont {A.}~\bibnamefont
  {Stern}},\ }\bibfield  {title} {\bibinfo {title} {Anyons and the quantum hall
  effect—a pedagogical review},\ }\href@noop {} {\bibfield  {journal}
  {\bibinfo  {journal} {Annals of Physics}\ }\textbf {\bibinfo {volume}
  {323}},\ \bibinfo {pages} {204} (\bibinfo {year} {2008})}\BibitemShut
  {NoStop}%
\bibitem [{\citenamefont {Lee}\ and\ \citenamefont
  {Fisher}(1989)}]{lee1989anyon}%
  \BibitemOpen
  \bibfield  {author} {\bibinfo {author} {\bibfnamefont {D.-H.}\ \bibnamefont
  {Lee}}\ and\ \bibinfo {author} {\bibfnamefont {M.~P.}\ \bibnamefont
  {Fisher}},\ }\bibfield  {title} {\bibinfo {title} {Anyon superconductivity
  and the fractional quantum hall effect},\ }\href@noop {} {\bibfield
  {journal} {\bibinfo  {journal} {Physical review letters}\ }\textbf {\bibinfo
  {volume} {63}},\ \bibinfo {pages} {903} (\bibinfo {year} {1989})}\BibitemShut
  {NoStop}%
\bibitem [{\citenamefont {Manna}\ \emph {et~al.}(2020)\citenamefont {Manna},
  \citenamefont {Pal}, \citenamefont {Wang}, \citenamefont {Nielsen} \emph
  {et~al.}}]{manna2020anyons}%
  \BibitemOpen
  \bibfield  {author} {\bibinfo {author} {\bibfnamefont {S.}~\bibnamefont
  {Manna}}, \bibinfo {author} {\bibfnamefont {B.}~\bibnamefont {Pal}}, \bibinfo
  {author} {\bibfnamefont {W.}~\bibnamefont {Wang}}, \bibinfo {author}
  {\bibfnamefont {A.~E.}\ \bibnamefont {Nielsen}}, \emph {et~al.},\ }\bibfield
  {title} {\bibinfo {title} {Anyons and fractional quantum hall effect in
  fractal dimensions},\ }\href@noop {} {\bibfield  {journal} {\bibinfo
  {journal} {Physical Review Research}\ }\textbf {\bibinfo {volume} {2}},\
  \bibinfo {pages} {023401} (\bibinfo {year} {2020})}\BibitemShut {NoStop}%
\bibitem [{\citenamefont {Tevosyan}\ and\ \citenamefont
  {MacDonald}(1997)}]{tevosyan1997virial}%
  \BibitemOpen
  \bibfield  {author} {\bibinfo {author} {\bibfnamefont {K.}~\bibnamefont
  {Tevosyan}}\ and\ \bibinfo {author} {\bibfnamefont {A.}~\bibnamefont
  {MacDonald}},\ }\bibfield  {title} {\bibinfo {title} {Virial expansions,
  exclusion statistics, and the fractional quantum hall effect},\ }\href@noop
  {} {\bibfield  {journal} {\bibinfo  {journal} {Physical Review B}\ }\textbf
  {\bibinfo {volume} {56}},\ \bibinfo {pages} {7517} (\bibinfo {year}
  {1997})}\BibitemShut {NoStop}%
\bibitem [{\citenamefont {Maleki}\ \emph
  {et~al.}(2023{\natexlab{a}})\citenamefont {Maleki}, \citenamefont
  {Mohammadzadeh}, \citenamefont {Najafi},\ and\ \citenamefont
  {Ebadi}}]{maleki2023beyond}%
  \BibitemOpen
  \bibfield  {author} {\bibinfo {author} {\bibfnamefont {M.}~\bibnamefont
  {Maleki}}, \bibinfo {author} {\bibfnamefont {H.}~\bibnamefont
  {Mohammadzadeh}}, \bibinfo {author} {\bibfnamefont {M.~N.}\ \bibnamefont
  {Najafi}},\ and\ \bibinfo {author} {\bibfnamefont {Z.}~\bibnamefont
  {Ebadi}},\ }\bibfield  {title} {\bibinfo {title} {Beyond the planck
  temperature},\ }\href@noop {} {\bibfield  {journal} {\bibinfo  {journal}
  {Physica Scripta}\ }\textbf {\bibinfo {volume} {98}},\ \bibinfo {pages}
  {045008} (\bibinfo {year} {2023}{\natexlab{a}})}\BibitemShut {NoStop}%
\bibitem [{\citenamefont {Maleki}\ \emph {et~al.}(2025)\citenamefont {Maleki},
  \citenamefont {Sattari},\ and\ \citenamefont
  {Mirershadi}}]{maleki2025deformed}%
  \BibitemOpen
  \bibfield  {author} {\bibinfo {author} {\bibfnamefont {M.}~\bibnamefont
  {Maleki}}, \bibinfo {author} {\bibfnamefont {F.}~\bibnamefont {Sattari}},\
  and\ \bibinfo {author} {\bibfnamefont {S.}~\bibnamefont {Mirershadi}},\
  }\bibfield  {title} {\bibinfo {title} {Deformed boson statistics and
  primordial black hole production under slow expansion conditions},\
  }\href@noop {} {\bibfield  {journal} {\bibinfo  {journal} {Chinese Journal of
  Physics}\ } (\bibinfo {year} {2025})}\BibitemShut {NoStop}%
\bibitem [{\citenamefont {Maleki}\ \emph
  {et~al.}(2023{\natexlab{b}})\citenamefont {Maleki}, \citenamefont {Ebadi},\
  and\ \citenamefont {Mohammadzadeh}}]{maleki2023q}%
  \BibitemOpen
  \bibfield  {author} {\bibinfo {author} {\bibfnamefont {M.}~\bibnamefont
  {Maleki}}, \bibinfo {author} {\bibfnamefont {Z.}~\bibnamefont {Ebadi}},\ and\
  \bibinfo {author} {\bibfnamefont {H.}~\bibnamefont {Mohammadzadeh}},\
  }\bibfield  {title} {\bibinfo {title} {q-deformed bose statistics and the
  gross pitaevskii equation},\ }\href@noop {} {\bibfield  {journal} {\bibinfo
  {journal} {International Journal of Geometric Methods in Modern Physics}\ }
  (\bibinfo {year} {2023}{\natexlab{b}})}\BibitemShut {NoStop}%
\bibitem [{\citenamefont {Maleki}\ \emph {et~al.}(2024)\citenamefont {Maleki},
  \citenamefont {Ebadi},\ and\ \citenamefont
  {Mohammadzadeh}}]{maleki2024nonextensive}%
  \BibitemOpen
  \bibfield  {author} {\bibinfo {author} {\bibfnamefont {M.}~\bibnamefont
  {Maleki}}, \bibinfo {author} {\bibfnamefont {Z.}~\bibnamefont {Ebadi}},\ and\
  \bibinfo {author} {\bibfnamefont {H.}~\bibnamefont {Mohammadzadeh}},\
  }\bibfield  {title} {\bibinfo {title} {Nonextensive statistics and the
  entropy on the horizon},\ }\href@noop {} {\bibfield  {journal} {\bibinfo
  {journal} {International Journal of Modern Physics A}\ }\textbf {\bibinfo
  {volume} {39}},\ \bibinfo {pages} {2450019} (\bibinfo {year}
  {2024})}\BibitemShut {NoStop}%
\bibitem [{\citenamefont {Salinas}\ and\ \citenamefont
  {Tsallis}(1999)}]{salinas1999special}%
  \BibitemOpen
  \bibfield  {author} {\bibinfo {author} {\bibfnamefont {S.~R.}\ \bibnamefont
  {Salinas}}\ and\ \bibinfo {author} {\bibfnamefont {C.}~\bibnamefont
  {Tsallis}},\ }\bibfield  {title} {\bibinfo {title} {Special issue on
  nonextensive statistical mechanics and thermodynamic},\ }\href@noop {}
  {\bibfield  {journal} {\bibinfo  {journal} {Braz. J. Phys}\ }\textbf
  {\bibinfo {volume} {29}} (\bibinfo {year} {1999})}\BibitemShut {NoStop}%
\bibitem [{\citenamefont {Thurner}\ and\ \citenamefont
  {Tsallis}(2005)}]{thurner2005nonextensive}%
  \BibitemOpen
  \bibfield  {author} {\bibinfo {author} {\bibfnamefont {S.}~\bibnamefont
  {Thurner}}\ and\ \bibinfo {author} {\bibfnamefont {C.}~\bibnamefont
  {Tsallis}},\ }\bibfield  {title} {\bibinfo {title} {Nonextensive aspects of
  self-organized scale-free gas-like networks},\ }\href@noop {} {\bibfield
  {journal} {\bibinfo  {journal} {Europhysics Letters}\ }\textbf {\bibinfo
  {volume} {72}},\ \bibinfo {pages} {197} (\bibinfo {year} {2005})}\BibitemShut
  {NoStop}%
\bibitem [{\citenamefont {Barboza~Jr}\ \emph {et~al.}(2015)\citenamefont
  {Barboza~Jr}, \citenamefont {Nunes}, \citenamefont {Abreu},\ and\
  \citenamefont {Neto}}]{barboza2015dark}%
  \BibitemOpen
  \bibfield  {author} {\bibinfo {author} {\bibfnamefont {E.~M.}\ \bibnamefont
  {Barboza~Jr}}, \bibinfo {author} {\bibfnamefont {R.~d.~C.}\ \bibnamefont
  {Nunes}}, \bibinfo {author} {\bibfnamefont {E.~M.}\ \bibnamefont {Abreu}},\
  and\ \bibinfo {author} {\bibfnamefont {J.~A.}\ \bibnamefont {Neto}},\
  }\bibfield  {title} {\bibinfo {title} {Dark energy models through
  nonextensive tsallis’ statistics},\ }\href@noop {} {\bibfield  {journal}
  {\bibinfo  {journal} {Physica A: Statistical Mechanics and its Applications}\
  }\textbf {\bibinfo {volume} {436}},\ \bibinfo {pages} {301} (\bibinfo {year}
  {2015})}\BibitemShut {NoStop}%
\bibitem [{\citenamefont {Beck}(2004)}]{beck2004nonextensive}%
  \BibitemOpen
  \bibfield  {author} {\bibinfo {author} {\bibfnamefont {C.}~\bibnamefont
  {Beck}},\ }\bibfield  {title} {\bibinfo {title} {Nonextensive scalar field
  theories and dark energy models},\ }\href@noop {} {\bibfield  {journal}
  {\bibinfo  {journal} {Physica A: Statistical Mechanics and its Applications}\
  }\textbf {\bibinfo {volume} {340}},\ \bibinfo {pages} {459} (\bibinfo {year}
  {2004})}\BibitemShut {NoStop}%
\bibitem [{\citenamefont {Leubner}(2005)}]{leubner2005nonextensive}%
  \BibitemOpen
  \bibfield  {author} {\bibinfo {author} {\bibfnamefont {M.}~\bibnamefont
  {Leubner}},\ }\bibfield  {title} {\bibinfo {title} {Nonextensive theory of
  dark matter and gas density profiles},\ }\href@noop {} {\bibfield  {journal}
  {\bibinfo  {journal} {The Astrophysical Journal}\ }\textbf {\bibinfo {volume}
  {632}},\ \bibinfo {pages} {L1} (\bibinfo {year} {2005})}\BibitemShut
  {NoStop}%
\bibitem [{\citenamefont {Zadeh}\ \emph {et~al.}(2018)\citenamefont {Zadeh},
  \citenamefont {Sheykhi}, \citenamefont {Moradpour},\ and\ \citenamefont
  {Bamba}}]{zadeh2018note}%
  \BibitemOpen
  \bibfield  {author} {\bibinfo {author} {\bibfnamefont {M.~A.}\ \bibnamefont
  {Zadeh}}, \bibinfo {author} {\bibfnamefont {A.}~\bibnamefont {Sheykhi}},
  \bibinfo {author} {\bibfnamefont {H.}~\bibnamefont {Moradpour}},\ and\
  \bibinfo {author} {\bibfnamefont {K.}~\bibnamefont {Bamba}},\ }\bibfield
  {title} {\bibinfo {title} {Note on tsallis holographic dark energy},\
  }\href@noop {} {\bibfield  {journal} {\bibinfo  {journal} {The European
  Physical Journal C}\ }\textbf {\bibinfo {volume} {78}},\ \bibinfo {pages} {1}
  (\bibinfo {year} {2018})}\BibitemShut {NoStop}%
\bibitem [{\citenamefont {Maleki}\ \emph
  {et~al.}(2023{\natexlab{c}})\citenamefont {Maleki}, \citenamefont
  {Mohammadzadeh},\ and\ \citenamefont {Ebadi}}]{maleki2023nonextensive}%
  \BibitemOpen
  \bibfield  {author} {\bibinfo {author} {\bibfnamefont {M.}~\bibnamefont
  {Maleki}}, \bibinfo {author} {\bibfnamefont {H.}~\bibnamefont
  {Mohammadzadeh}},\ and\ \bibinfo {author} {\bibfnamefont {Z.}~\bibnamefont
  {Ebadi}},\ }\bibfield  {title} {\bibinfo {title} {Nonextensive gross
  pitaevskii equation},\ }\href@noop {} {\bibfield  {journal} {\bibinfo
  {journal} {International Journal of Geometric Methods in Modern Physics}\
  }\textbf {\bibinfo {volume} {20}},\ \bibinfo {pages} {2350216} (\bibinfo
  {year} {2023}{\natexlab{c}})}\BibitemShut {NoStop}%
\bibitem [{\citenamefont {Ebadi}\ \emph {et~al.}(2013)\citenamefont {Ebadi},
  \citenamefont {Mirza},\ and\ \citenamefont
  {Mohammadzadeh}}]{ebadi2013infinite}%
  \BibitemOpen
  \bibfield  {author} {\bibinfo {author} {\bibfnamefont {Z.}~\bibnamefont
  {Ebadi}}, \bibinfo {author} {\bibfnamefont {B.}~\bibnamefont {Mirza}},\ and\
  \bibinfo {author} {\bibfnamefont {H.}~\bibnamefont {Mohammadzadeh}},\
  }\bibfield  {title} {\bibinfo {title} {Infinite statistics condensate as a
  model of dark matter},\ }\href@noop {} {\bibfield  {journal} {\bibinfo
  {journal} {Journal of Cosmology and Astroparticle Physics}\ }\textbf
  {\bibinfo {volume} {2013}}\bibinfo  {number} { (11)},\ \bibinfo {pages}
  {057}}\BibitemShut {NoStop}%
\bibitem [{\citenamefont {Gavrilik}\ \emph {et~al.}(2018)\citenamefont
  {Gavrilik}, \citenamefont {Kachurik}, \citenamefont {Khelashvili},\ and\
  \citenamefont {Nazarenko}}]{gavrilik2018condensate}%
  \BibitemOpen
\bibfield  {number} {  }\bibfield  {author} {\bibinfo {author} {\bibfnamefont
  {A.}~\bibnamefont {Gavrilik}}, \bibinfo {author} {\bibfnamefont
  {I.}~\bibnamefont {Kachurik}}, \bibinfo {author} {\bibfnamefont
  {M.}~\bibnamefont {Khelashvili}},\ and\ \bibinfo {author} {\bibfnamefont
  {A.}~\bibnamefont {Nazarenko}},\ }\bibfield  {title} {\bibinfo {title}
  {Condensate of $\mu$-bose gas as a model of dark matter},\ }\href@noop {}
  {\bibfield  {journal} {\bibinfo  {journal} {Physica A: Statistical Mechanics
  and its Applications}\ }\textbf {\bibinfo {volume} {506}},\ \bibinfo {pages}
  {835} (\bibinfo {year} {2018})}\BibitemShut {NoStop}%
\bibitem [{\citenamefont {Maleki}\ \emph {et~al.}(2019)\citenamefont {Maleki},
  \citenamefont {Mohammadzadeh}, \citenamefont {Ebadi},\ and\ \citenamefont
  {Najafi}}]{maleki2019deformed}%
  \BibitemOpen
  \bibfield  {author} {\bibinfo {author} {\bibfnamefont {M.}~\bibnamefont
  {Maleki}}, \bibinfo {author} {\bibfnamefont {H.}~\bibnamefont
  {Mohammadzadeh}}, \bibinfo {author} {\bibfnamefont {Z.}~\bibnamefont
  {Ebadi}},\ and\ \bibinfo {author} {\bibfnamefont {M.~N.}\ \bibnamefont
  {Najafi}},\ }\bibfield  {title} {\bibinfo {title} {Deformed boson condensate
  as a model of dark matter},\ }\href@noop {} {\bibfield  {journal} {\bibinfo
  {journal} {arXiv preprint arXiv:1912.04656}\ } (\bibinfo {year}
  {2019})}\BibitemShut {NoStop}%
\bibitem [{\citenamefont {Maleki}\ \emph
  {et~al.}(2022{\natexlab{a}})\citenamefont {Maleki}, \citenamefont
  {Mohammadzadeh}, \citenamefont {Ebadi},\ and\ \citenamefont
  {Najafi}}]{maleki2022description}%
  \BibitemOpen
  \bibfield  {author} {\bibinfo {author} {\bibfnamefont {M.}~\bibnamefont
  {Maleki}}, \bibinfo {author} {\bibfnamefont {H.}~\bibnamefont
  {Mohammadzadeh}}, \bibinfo {author} {\bibfnamefont {Z.}~\bibnamefont
  {Ebadi}},\ and\ \bibinfo {author} {\bibfnamefont {M.~N.}\ \bibnamefont
  {Najafi}},\ }\bibfield  {title} {\bibinfo {title} {Description of the
  properties of dark matter by deformed statistics},\ }\href@noop {} {\bibfield
   {journal} {\bibinfo  {journal} {Journal of Statistical Mechanics: Theory and
  Experiment}\ }\textbf {\bibinfo {volume} {2022}},\ \bibinfo {pages} {013104}
  (\bibinfo {year} {2022}{\natexlab{a}})}\BibitemShut {NoStop}%
\bibitem [{\citenamefont {Maleki}\ \emph
  {et~al.}(2022{\natexlab{b}})\citenamefont {Maleki}, \citenamefont
  {Mohammadzadeh}, \citenamefont {Ebadi},\ and\ \citenamefont
  {Najafi}}]{maleki2022mass}%
  \BibitemOpen
  \bibfield  {author} {\bibinfo {author} {\bibfnamefont {M.}~\bibnamefont
  {Maleki}}, \bibinfo {author} {\bibfnamefont {H.}~\bibnamefont
  {Mohammadzadeh}}, \bibinfo {author} {\bibfnamefont {Z.}~\bibnamefont
  {Ebadi}},\ and\ \bibinfo {author} {\bibfnamefont {M.~N.}\ \bibnamefont
  {Najafi}},\ }\bibfield  {title} {\bibinfo {title} {Mass bounds on the
  deformed bosons as dark matter particles},\ }\href@noop {} {\bibfield
  {journal} {\bibinfo  {journal} {International Journal of Modern Physics A}\
  }\textbf {\bibinfo {volume} {37}},\ \bibinfo {pages} {2250059} (\bibinfo
  {year} {2022}{\natexlab{b}})}\BibitemShut {NoStop}%
\bibitem [{\citenamefont {Biedenharn}(1989)}]{biedenharn1989quantum}%
  \BibitemOpen
  \bibfield  {author} {\bibinfo {author} {\bibfnamefont {L.}~\bibnamefont
  {Biedenharn}},\ }\bibfield  {title} {\bibinfo {title} {The quantum group suq
  (2) and a q-analogue of the boson operators},\ }\href@noop {} {\bibfield
  {journal} {\bibinfo  {journal} {Journal of Physics A: Mathematical and
  General}\ }\textbf {\bibinfo {volume} {22}},\ \bibinfo {pages} {L873}
  (\bibinfo {year} {1989})}\BibitemShut {NoStop}%
\bibitem [{\citenamefont {Macfarlane}(1989)}]{macfarlane1989q}%
  \BibitemOpen
  \bibfield  {author} {\bibinfo {author} {\bibfnamefont {A.}~\bibnamefont
  {Macfarlane}},\ }\bibfield  {title} {\bibinfo {title} {On q-analogues of the
  quantum harmonic oscillator and the quantum group su (2) q},\ }\href@noop {}
  {\bibfield  {journal} {\bibinfo  {journal} {Journal of Physics A:
  Mathematical and general}\ }\textbf {\bibinfo {volume} {22}},\ \bibinfo
  {pages} {4581} (\bibinfo {year} {1989})}\BibitemShut {NoStop}%
\bibitem [{\citenamefont {Mirza}\ and\ \citenamefont
  {Mohammadzadeh}(2011)}]{mirza2011thermodynamic}%
  \BibitemOpen
  \bibfield  {author} {\bibinfo {author} {\bibfnamefont {B.}~\bibnamefont
  {Mirza}}\ and\ \bibinfo {author} {\bibfnamefont {H.}~\bibnamefont
  {Mohammadzadeh}},\ }\bibfield  {title} {\bibinfo {title} {Thermodynamic
  geometry of deformed bosons and fermions},\ }\href@noop {} {\bibfield
  {journal} {\bibinfo  {journal} {Journal of Physics A: Mathematical and
  Theoretical}\ }\textbf {\bibinfo {volume} {44}},\ \bibinfo {pages} {475003}
  (\bibinfo {year} {2011})}\BibitemShut {NoStop}%
\bibitem [{\citenamefont {Oh}\ \emph {et~al.}(2011)\citenamefont {Oh},
  \citenamefont {de~Blok}, \citenamefont {Brinks}, \citenamefont {Walter},\
  and\ \citenamefont {Kennicutt}}]{oh2011dark}%
  \BibitemOpen
  \bibfield  {author} {\bibinfo {author} {\bibfnamefont {S.-H.}\ \bibnamefont
  {Oh}}, \bibinfo {author} {\bibfnamefont {W.~J.~G.}\ \bibnamefont {de~Blok}},
  \bibinfo {author} {\bibfnamefont {E.}~\bibnamefont {Brinks}}, \bibinfo
  {author} {\bibfnamefont {F.}~\bibnamefont {Walter}},\ and\ \bibinfo {author}
  {\bibfnamefont {R.~C.}\ \bibnamefont {Kennicutt}, \bibfnamefont {Jr.}},\
  }\bibfield  {title} {\bibinfo {title} {Dark and luminous matter in things
  dwarf galaxies},\ }\href@noop {} {\bibfield  {journal} {\bibinfo  {journal}
  {The Astronomical Journal}\ }\textbf {\bibinfo {volume} {141}},\ \bibinfo
  {pages} {193} (\bibinfo {year} {2011})},\ \bibinfo {note}
  {arXiv:1011.0899}\BibitemShut {NoStop}%
\bibitem [{\citenamefont {Swaters}\ \emph {et~al.}(2003)\citenamefont
  {Swaters}, \citenamefont {Verheijen}, \citenamefont {Bershady},\ and\
  \citenamefont {Andersen}}]{swaters2003central}%
  \BibitemOpen
  \bibfield  {author} {\bibinfo {author} {\bibfnamefont {R.~A.}\ \bibnamefont
  {Swaters}}, \bibinfo {author} {\bibfnamefont {M.~A.~W.}\ \bibnamefont
  {Verheijen}}, \bibinfo {author} {\bibfnamefont {M.~A.}\ \bibnamefont
  {Bershady}},\ and\ \bibinfo {author} {\bibfnamefont {D.~R.}\ \bibnamefont
  {Andersen}},\ }\bibfield  {title} {\bibinfo {title} {The central mass
  distribution in dwarf and low surface brightness galaxies},\ }\href@noop {}
  {\bibfield  {journal} {\bibinfo  {journal} {The Astrophysical Journal}\
  }\textbf {\bibinfo {volume} {587}},\ \bibinfo {pages} {L19} (\bibinfo {year}
  {2003})}\BibitemShut {NoStop}%
\bibitem [{\citenamefont {Carignan}\ and\ \citenamefont
  {Beaulieu}(1989)}]{carignan1989optical}%
  \BibitemOpen
  \bibfield  {author} {\bibinfo {author} {\bibfnamefont {C.}~\bibnamefont
  {Carignan}}\ and\ \bibinfo {author} {\bibfnamefont {S.}~\bibnamefont
  {Beaulieu}},\ }\bibfield  {title} {\bibinfo {title} {Optical and h i studies
  of the gas-rich dwarf irregular galaxy ddo 154},\ }\href@noop {} {\bibfield
  {journal} {\bibinfo  {journal} {The Astrophysical Journal}\ }\textbf
  {\bibinfo {volume} {347}},\ \bibinfo {pages} {760} (\bibinfo {year}
  {1989})}\BibitemShut {NoStop}%
\end{thebibliography}%

\end{document}